\documentclass{aa}  

\usepackage{graphicx}
\usepackage{txfonts}
\usepackage{lipsum}
\usepackage{subcaption}         % necessary for continued figures, example in section 3
\usepackage{lscape}             % to rotate a single page table, example in appendix.
\usepackage{placeins}           % useful with \FloatBarrier, to keep 
\usepackage[table,xcdraw]{xcolor}        
\usepackage[breaklinks, colorlinks, citecolor=blue, urlcolor = blue]{hyperref}
\usepackage{multirow}
\usepackage{aas_macros}
\begin{document}

%%%%%%%%%%%%%%%%%%%%%%%%%%%%%%%%%%%%%%%%
% if you use custom commands in your title,
% ensure to check your title when submitting!
%%%%%%%%%%%%%%%%%%%%%%%%%%%%%%%%%%%%%%%%
   \title{Symmetric kiloparsec-scale radio knots in NGC 7213}

   \subtitle{Evidence of a confined weak jet and recurrent nuclear activity}

%%%%%%%%%%%%%%%%%%%%%%%%%%%%%%%%%%%%%%%%
% Please separate each author with the \and command
%
% Please do not include ORCIDs next to author names.
% Only ORCIDs authenticated by individual authors in EDPS
% editorial system will be taken into account.
% ORCIDs included here will be removed.
%%%%%%%%%%%%%%%%%%%%%%%%%%%%%%%%%%%%%%%%

   \author{ G. Bruni\inst{1}\and 
            F. Panessa\inst{1} \and 
            E. Kammoun\inst{2} \and
            A. L. Thakur\inst{1} \and
            C. Reynolds\inst{3} \and
            R. Ricci\inst{4,5} \and
            M. Wieringa\inst{6} \and
            S. Bianchi\inst{7} \and\\
            A. De Rosa\inst{1} \and
            G. Matt\inst{7} \and
            F. Nicastro\inst{8} \and
            F. Ursini\inst{7}
        }

   \institute{INAF -- Istituto di Astrofisica e Planetologia Spaziali, Via Fosso del Cavaliere, 00133 Roma, Italy\\
             \email{gabriele.bruni@inaf.it}
             %\thanks{Shows the usage of elements in the institute field}
            \and Cahill Center for Astronomy \& Astrophysics, California Institute of Technology, 1216 East California Boulevard, Pasadena, CA 91125, USA
            \and CSIRO Astronomy and Space Science, P.O. Box 1130, Bentley WA 6102, Australia
            \and INAF -- Istituto di Radioastronomia, via Gobetti 101, 40129 Bologna, Italy 
            \and CSIRO Astronomy and Space Science, PO Box 76, Epping, NewSouth Wales 1710, Australia
            \and Dipartimento di Fisica, Universit\'a di Roma Tor Vergata, via della Ricerca Scientifica 1, 00133 Roma, Italy 
            \and Dipartimento di Matematica e Fisica, Università degli Studi Roma Tre, Via della Vasca Navale 84, I-00146, Roma, Italy
            \and INAF -- Osservatorio Astronomico di Roma, Via Frascati 33, 00040 Monte Porzio Catone, Roma, Italy 
            \\ }

   \date{Received September 30, 20XX}

% \abstract{}{}{}{}{}
% 5 {} token are mandatory
 
\abstract
% Context
{Low-luminosity active galactic nuclei (LLAGNs) often host weak radio jets whose propagation and impact on their host galaxies are strongly influenced by the surrounding interstellar medium. Constraining the connection between nuclear activity, radio variability, and large-scale radio structures in these systems remains a key observational challenge.}
% Aims
{We investigate the radio properties of the nearby LLAGN NGC~7213 to assess its ability to launch collimated outflows beyond the nuclear region and to characterise the nature, origin, and variability of newly identified radio components on parsec to kiloparsec scales.}
% Methods
{We present new radio observations obtained with MeerKAT, uGMRT, ATCA, and the Australian Long Baseline Array (LBA), covering frequencies from 300~MHz to 9~GHz. We analysed the morphology and spectra of the radio emission on kiloparsec scales and performed a dedicated LBA monitoring campaign to probe the parsec-scale core and its variability.}
% Results
{We discovered a pair of compact radio knots located symmetrically at a projected distance of $\sim$5~kpc north and south of the nucleus of NGC~7213. The two knots exhibit nearly identical flux densities and flat radio spectra from 300~MHz up to at least 5.5~GHz, with no significant spectral or geometric asymmetry. ATCA observations confirm the flat spectra at higher frequencies. The LBA monitoring shows that the nuclear radio source remains unresolved at all epochs, constraining the 8~GHz emission to sub-parsec scales. Significant variability is detected on both decade-long timescales, relative to archival LBA data, and on month-long timescales during our monitoring, with a flux-density increase of $\sim$40~mJy over six months.}
% Conclusions
{The symmetry, spectra, and physical properties of the kiloparsec-scale knots strongly support their interpretation as compact termination shocks of a weak or intermittent jet launched by the LLAGN in NGC~7213 and confined by the dense, disturbed interstellar medium of the host galaxy. The unresolved yet variable parsec-scale core traced by the LBA indicates that the observed high-frequency radio variability originates in the innermost jet region, which is likely linked to the recent increase in nuclear activity. NGC~7213 thus provides a nearby example of how weak jets in low-accretion AGN can produce both compact nuclear variability and symmetric kiloparsec-scale structures when interacting with complex environments.}


    \maketitle
    \nolinenumbers

%%%%%%%%%%%%%%%%%%%%%%%%%%%%%%%%%%%%%%%%%%%%%%%%%%%%%%%%%%%%%%
\section{Introduction}

NGC\,7213 is a nearby galaxy hosting one of the closest known
low-luminosity active galactic nuclei (LLAGN), at a distance of 
$D\simeq22$--23\,Mpc \citep{Tully1988}.  
Its nucleus displays the combined spectroscopic characteristics
of a Seyfert~1 and a low-ionisation nuclear emission-line region (LINER;
\citealt{Phillips1979,Filippenko1984}), 
with broad permitted lines, a strong blue continuum, and prominent
low-ionisation narrow emission lines.  
Despite its modest bolometric luminosity
($L_{\rm bol} \simeq 1.7\times10^{43}$\,erg\,s$^{-1}$;
\citealt{Emmanoulopoulos2012}),
NGC\,7213 has attracted considerable interest across multiple 
wavelengths owing to its peculiar accretion properties, complex
circumnuclear gas distribution, and evidence of episodic outflows.

Over the past decades, radio and X-ray observations have established
NGC\,7213 as a low-accretion system powered by a supermassive black hole
of mass $M_{\rm BH} = 8^{+16}_{-6}\times10^{7}\,M_\odot$
\citep{SchnorrMueller2014} accreting at a low Eddington ratio
\citep{Emmanoulopoulos2012,Ursini2015}.
The source shows a `harder-when-brighter' X-ray behaviour and
correlated X-ray/radio variability, with radio flares lagging the
X-rays by a few tens of days, suggestive of a weak compact jet
analogous to the hard state of X-ray binaries
\citep{Emmanoulopoulos2012,Bell2011}.  
Previous X-ray studies with \textit{Chandra}, \textit{XMM--Newton},
and \textit{NuSTAR} revealed a power-law continuum with a high-energy
cut-off but no obvious Compton reflection hump, plus a narrow
Fe\,K$\alpha$ line and weaker Fe\,\textsc{xxv}/Fe\,\textsc{xxvi}
features, pointing to reprocessing in Compton-thin material with a
low covering factor rather than in a classical, Compton-thick torus
\citep{Bianchi2003,Lobban2010,Ursini2015}.

In the optical band, NGC\,7213 exhibits a remarkably complex broad-line
region (BLR).  
In addition to a `classical' symmetric broad H$\alpha$ component,
the source shows a long-lived, double-peaked Balmer profile
\citep{SchnorrMueller2014,Schimoia2017}, placing it within the class of 
AGNs well known for exhibiting double-peaked Balmer emission lines
\citep[e.g.][]{EracleousHalpern1994,StorchiBergmann2017}.  
Recent work has significantly expanded our understanding of this
population:  
\citet{Ward2025} used SDSS-V and high-cadence optical spectroscopy to 
demonstrate that double-peaked emitters form a heterogeneous class 
whose line shapes, variability, and Balmer decrement are tightly connected 
to the structure of the outer accretion disk and to disk asymmetries such 
as spiral arms or eccentric annuli.  
These results reinforce the interpretation of the H$\alpha$ profile of 
NGC\,7213 as a disk-origin emission, consistent with dynamical modelling of 
its variable double-peaked component \citep{Schimoia2017} and with the idea 
that at least part of the BLR directly traces the outer accretion disk.

This global picture has been significantly refined by the recent
XRISM/Resolve campaign of NGC\,7213, supported by simultaneous
\textit{XMM--Newton}, \textit{NuSTAR}, and SOAR optical spectroscopy
\citep{Kammoun2025}.  
The high-resolution Resolve spectrum resolves the neutral
Fe\,K$\alpha$ into two components: (i) a narrow core with full width half maximum (FWHM) $\sim 650$~km\,s$^{-1}$ produced at radii consistent with the
dust sublimation region, and (ii) a broader, asymmetric component
modelled as relativistic disk emission from
$R_{\rm in}\sim 10^{2}\,R_{\rm g}$, viewed at a low inclination
$i\sim 10^\circ$.  
In addition, broadened Fe\,\textsc{xxv} and Fe\,\textsc{xxvi} lines
trace gas located between the inner disk and the optical BLR.
Together with the weak Compton hump and the small equivalent width of
the narrow Fe\,K$\alpha$ core, these results reveal a radially and
vertically stratified accretion structure extending over 
$\sim$10$^{2}$--10$^{5}\,R_{\rm g}$ and place NGC\,7213 in an
intermediate-accretion regime where the inner thin disk and BLR are
present, while the classical torus has a low covering fraction.

On larger scales, NGC\,7213 exhibits clear signatures of past
dynamical disturbance.  
\citet{Hameed2001} discovered a $\sim 19$\,kpc-long H$\alpha$
filament located $\sim 18.6$\,kpc south of the nucleus and spatially
coincident with an H\,\textsc{i} tidal tail, revealing a disturbed and
asymmetric neutral-gas distribution likely caused by a past merger.
The filament has been interpreted either as gas photo-ionised by the 
AGN or shock-ionised by a previous episode of jet activity, suggesting 
that NGC\,7213 may have undergone earlier radio-loud phases no longer 
visible in contemporary GHz data.

Despite this wealth of multi-wavelength information, the radio
structure of NGC\,7213 on kiloparsec scales has remained poorly
constrained, with only a compact, flat-spectrum VLBI core known prior 
to this work \citep{Blank2005,Bell2011}.  
Whether the AGN in NGC\,7213 is capable of driving jets beyond the 
central few hundred parsecs has thus remained an open question.

In this work we present new MeerKAT, uGMRT, ATCA, and LBA observations that reveal
for the first time a pair of symmetric radio knots located at
$\sim 5$\,kpc north and south of the nucleus.  
We interpret these knots as kiloparsec-scale hotspots tracing
collimated outflows from the AGN, and we discuss their morphology,
spectral properties, and environment in the context of recurrent or
intermittently powered jet activity and of the stratified accretion
flow inferred from recent high-resolution X-ray spectroscopy.

%%%%%%%%%%%%%%%%%%%%%%%%%%%%%%%%%%%%%%%%%%%%%%%%%%%%%%%%%%%%%%

\section{Observations and data reduction}

In the following, we present the different data collected in this work. An observations journal is given in Table\,\ref{tab:uGMRT} for uGMRT and MeerKAT, Table\,\ref{tab:LBA} for the LBA monitoring, and Table C.1 and D.1 (available at the CDS) for ATCA and \textit{Swift}.

\subsection{MeerKAT}
We made use of archival MeerKAT observations (project ID SCI-20220822-LM-01;
PI: Cotton) obtained in January 2023 in the L band (856--1712~MHz).
Raw visibilities were reduced using the \textsc{OxKAT} pipeline
\citep{Heywood2020}, a semi-automated calibration and imaging framework
developed for MeerKAT continuum observations. The pipeline performs
automated radio-frequency interference flagging, delay and bandpass
calibration, complex gain calibration, and flux-density scale
calibration using standard calibrator models. Imaging was carried out
with \textsc{WSCLEAN} \citep{Offringa2014,Offringa2017}, adopting
multi-frequency synthesis and multi-scale deconvolution to accurately
reconstruct both compact and extended emission. Briggs weighting was
used to achieve an optimal balance between sensitivity and angular
resolution. Several rounds of phase-only and amplitude self-calibration
were applied to improve the image fidelity. The final images reach rms
noise levels close to the theoretical thermal limit.

\subsection{uGMRT}
\label{sec:ugmrt}

The uGMRT observations were carried out in Band~3 (250--500~MHz) and
Band~4 (550--850~MHz) in November 2025 (Project ID 49\_074, PI Bruni), for a total of 4 hours in each band.
Data were reduced using the Source Peeling and Atmospheric
Modeling (\textsc{SPAM}) pipeline \citep{Intema2014}, which is optimised
for low-frequency interferometric observations. The pipeline performs
automated flagging of radio-frequency interference, initial calibration
of delays, bandpass, and complex gains using standard flux-density
calibrators, and applies an ionospheric calibration based on direction-dependent phase solutions. Imaging was performed using \textsc{WSCLEAN} \citep{Offringa2014,Offringa2017},
adopting wide-field imaging and multi-scale deconvolution to accurately
reconstruct both compact and extended emission.

\subsection{ATCA}
\label{sec:atca}

A 6-month ATCA monitoring was realised under DDT project CX579 (PI: Panessa), from July 2024 to February 2025.
Observations used the 4\,cm receiver, covering
the C and X bands (4.5--10.0~GHz). Each epoch consisted of a 1-hour run, of which $\sim$30 minutes were on-source. Flux-density calibration was performed using the standard ATCA primary
calibrator PKS~B1934--638.
Visibilities were calibrated and imaged using the
\textsc{MIRIAD} software package \citep{Sault1995}, following standard
procedures for continuum data. After flagging of radio-frequency
interference, bandpass and complex gain calibration were performed using
primary and secondary calibrators. Imaging was carried out using
multi-frequency synthesis, and deconvolution was performed with CLEAN.
Several rounds of phase-only self-calibration were applied to improve
the image fidelity. Flux densities were placed on the standard ATCA
flux-density scale, and the final images reach rms noise levels
consistent with the achieved sensitivity.

The flux densities collected with the monitoring are given in Appendix \ref{app:ATCA}: The first epoch was used to produce an image at 5.5 GHz, while at 9 GHz a co-added image of all data collected in configuration 6A was produced in order to maximise the angular resolution and sensitivity. 

\subsection{LBA}
\label{sec:lba}
The Long Baseline Array (LBA) observations were performed in X band, centred at 8.4~GHz, with
a total bandwidth of 256~MHz. A total of 4 epochs were collected (see table \ref{tab:LBA}), under project V657 (PI: Panessa).
The LBA observations were reduced following
standard VLBI procedures using the \textsc{AIPS} software package
\citep{Greisen2003}. Initial data inspection and flagging were performed
to remove corrupted data and radio-frequency interference. Amplitude
calibration was carried out using measured system temperatures and
antenna gain curves, while delay and phase calibration were obtained
from observations of fringe-finder and phase-calibrator sources. Fringe
fitting was applied to correct for residual delays and rates.

Imaging and self-calibration were performed using \textsc{DIFMAP}
\citep{Shepherd1997}. The data were iteratively imaged and self-
calibrated, starting with phase-only self-calibration and, where
appropriate, followed by amplitude self-calibration. The final images
were produced using CLEAN deconvolution and reach rms noise levels
consistent with the expected thermal sensitivity. The error budget is thus dominated by a 15\% uncertainty on the flux scale. At all epochs, the
source remains unresolved at the angular resolution of the LBA.

\subsection{Swift/XRT}
\label{sec:swift}
\begin{figure*}
\sidecaption
    \includegraphics[width=12cm]{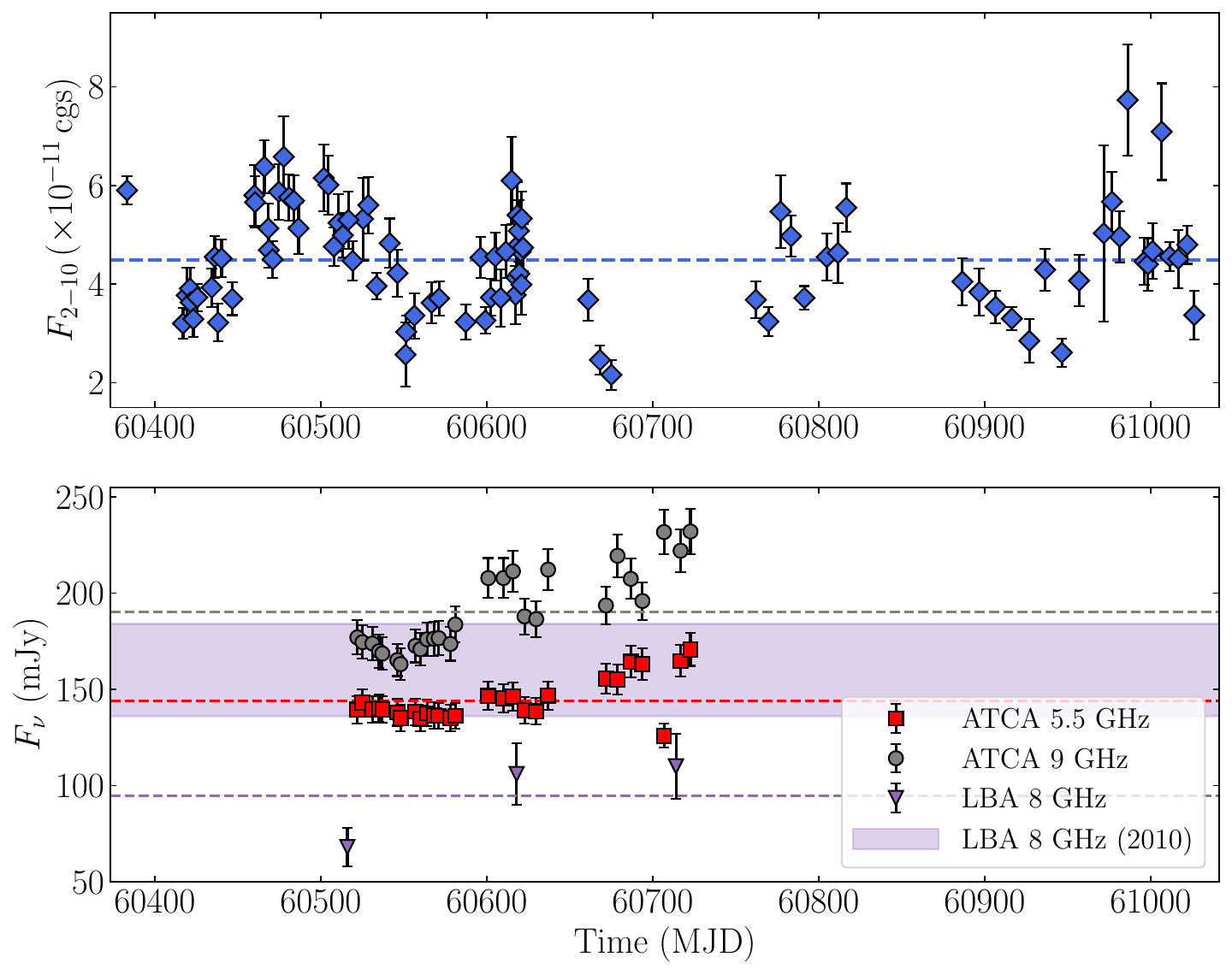}
    \caption{NGC\,7213 X-ray and radio light curves. Top panel: \textit{Swift}/XRT 2--10\,keV  light curve. Bottom panel: Radio light curves obtained from ATCA 5.5 GHz and 9\,GHz (red squares and grey circles, respectively), and LBA 8\,GHz (purple triangles). The purple shaded area shows the archival LBA 8\,GHz archival flux density measured on 2010 March 11. Dashed horizontal lines show the average fluxes corresponding to each of the light curves.}
    \label{fig:LBA_variability}
\end{figure*}

The X-ray Telscope (XRT) on-board the \textit{Neil Gehrels Swift Observatory} (\textit{Swift})  performed 79 observations of NGC\,7213 between 2024-03-14 and 2025-12-17, for a total of 80\,ks. We extracted the spectra from each of the observations using the \textit{Swift}/XRT automatic data product generator\footnote{\url{ https://www.Swift.ac.uk/user_objects/}} \citep{Evans2009}. We re-binned all the spectra using the optimal binning scheme in FTOOLS \citep{Kaastra2016}, with a minimum signal-to-noise ratio of 4. We then fitted each of the spectra (in the $0.4-8$\,keV range) with an absorbed power-law model considering Galactic absorption only \citep[$N_{\rm H} =1.08 \times 10^{20}\,\rm cm^{-2}$;][]{HI4PI}. The best-fit results are shown in Table D.1 (available at the CDS). The top panel of Fig.\,\ref{fig:LBA_variability} shows the $2-10$\,keV light curve during this period.

%%%%%%%%%%%%%%%%%%%%%%%%%%%%%%%%%%%%%%%%%%%%%%%%%%%%%%%%%%%%%%

%%%%%%%%%%%%%%%%%%%%%%%%%%%%%%%%%%%%%%%%%%%%%%%%%%%%%%%%%%%%%%
%%%%%%%%%%%%%%%%%%%%%%%%%%%%%%%%%%%%%%%%%%%%%%%%%%

%%%%%%%%%%%%%%%%%%%%%%%%%%%%%%%%%%%%%%%%%%%%%%%%%%
%%%%%%%%%%%%%%%%%%%%%%%%%%%%%%%%%%%%%%%%%%%%%%%%%%%%%%%%%%%%%%

\section{The radio/X-ray monitoring}

Triggered by the X-ray brightening of NGC\,7213 reported by \citet{Kammoun2025}, we initiated a coordinated multi-wavelength campaign to probe the connection between nuclear activity, parsec-scale radio variability, and the newly discovered kiloparsec-scale structures. The programme combined a dedicated 6-month LBA monitoring at 8\,GHz (probing the unresolved core on sub-parsec scales) with simultaneous ATCA observations at 5.5 and 9\,GHz and Swift/XRT pointings in the 2--10\,keV band (see Appendix \ref{app:ATCA} and \ref{app:Swift}, respectively).

\subsection{ATCA and Swift/XRT}
\label{sec:atca_swift}
We quantified the intrinsic variability of the ATCA light curves by means of the normalised excess variance \(\sigma^2_{\rm NXS}\) \citep{Vaughan2003}, from which the fractional variability amplitude is derived as
\[
F_{\rm var} = \sqrt{\frac{\sigma^2_{\rm NXS} - \sigma^2_{\rm err}}{\langle F \rangle^2}},
\]
where \(\sigma^2_{\rm err}\) is the mean squared measurement error and \(\langle F \rangle\) is the mean flux density. This yields \(F_{\rm var}=0.058\pm0.011\) at 5.5\,GHz and \(F_{\rm var}=0.100\pm0.010\) at 9.0\,GHz, indicating that the source is significantly more variable at the higher frequency.

The same formalism applied to the Swift/XRT 2--10\,keV light curve gives \(F_{\rm var}=0.27\pm0.02\). This value is substantially larger than those measured in the radio band, confirming that the variability amplitude decreases from the X-ray to the radio regime, as expected in a disk--jet coupling scenario.
Given the different cadences and the relatively sparse sampling of the simultaneous radio and X-ray monitoring, it was not possible to perform a meaningful cross-correlation analysis or to search for radio/X-ray lags with the present dataset.

Finally, to test for long-term evolution during the monitoring, we fitted the radio light curves with a weighted linear model. We find a positive slope at both frequencies, with \(dS/dt = (0.096\pm0.023)\) mJy d$^{-1}$ at 5.5\,GHz and \(dS/dt = (0.282\pm0.030)\) mJy d$^{-1}$ at 9.0\,GHz, showing that the flux increase is particularly significant at 9.0\,GHz.
Using the mean ATCA flux density at 5.5\,GHz (\(\langle S_{5.5} \rangle = 144\pm 3\) mJy, representative of the nuclear emission over the 27 epochs of Table~B.1) and the mean Swift/XRT 2--10\,keV flux during the same period, we computed the radio-loudness parameter
\[
R_X = \log\left(\frac{L_{5\,\rm GHz}}{L_{2-10\,keV}}\right),
\]
where \(L_{5\,\rm GHz} = 4\pi D^2\,\nu\,S_\nu\) (with \(\nu = 5\) GHz) and \(L_{2-10\,keV}\) is the X-ray luminosity corrected for Galactic absorption \citep{Terashima2003}. We obtain \( R_X = -4.2 \pm 0.1 \), consistent with the typical range observed in radio-quiet AGN (\( R_X < -4 \)). However, this value lies within the radio-loud regime when considering local low-luminosity AGN \citep{Panessa2007}.

To explore whether the radio-loudness itself varies during the campaign, we divided the 27 ATCA epochs into three roughly equal time intervals of 9 epochs each:  
(i) early phase (MJD 60521--60559),  
(ii) intermediate phase (MJD 60563--60622),  
(iii) late phase (MJD 60629--60722).  
Recomputing \(R_X\) in each bin using the corresponding mean radio and X-ray fluxes yields values consistent with each other within the uncertainties (\(\Delta R_X \lesssim 0.3\)), indicating that the radio-loudness parameter remained stable over the six-month period despite the observed flux increase in both bands. The results of the variability analysis are summarised in Table~\ref{tab:variability}.

%%%%%%%%%%%%%%%%%%%%%%%%%%%%%%%%%%%%%%%%%%%
\begin{table*}
\caption{Summary of variability parameters derived from the ATCA radio and Swift/XRT X-ray monitoring of NGC\,7213.}
\label{tab:variability}
\centering
\begin{tabular}{lccc}
\hline\hline
Parameter & Value & Frequency / Band & Note \\
\hline
\multirow{3}{*}{$F_{\rm var}$} 
 & $0.058 \pm 0.011$ & 5.5\,GHz & ATCA \\
 & $0.100 \pm 0.010$ & 9.0\,GHz  & ATCA \\
 & $0.27 \pm 0.02$   & 2--10\,keV & Swift/XRT \\
\hline
\multirow{2}{*}{$dS/dt$ (mJy\,d$^{-1}$)} 
 & $+0.096 \pm 0.023$ & 5.5\,GHz & linear trend \\
 & $+0.282 \pm 0.030$ & 9.0\,GHz  & linear trend \\
\hline
\end{tabular}
\end{table*}

%%%%%%%%%%%%%%%%%%%%%%%%%%%%%%%%%%%%%%%%%%%

\subsection{Variability on the pc-scale with LBA}
\label{sec:lba_monitoring}

At all observed epochs, including the final one obtained in February 2025, the source remains unresolved at the angular resolution of the LBA. The restored beams range from $\sim2.0\times1.3$ to $\sim3.1\times1.9$ mas, corresponding to physical scales of $\lesssim0.2$--0.3~pc at the distance of NGC~7213. This implies that the bulk of the 8~GHz emission originates within the innermost parsec-scale region of the AGN. While no resolved jet-like structure is detected, the non-detection of extended emission does not rule out the presence of a compact or newly emerging component below the current resolution limit, which could become detectable at future epochs if it expands or brightens.

Given the limited number of LBA epochs (only three during the monitoring campaign), a statistical variability analysis based on the normalised excess variance could not be performed, unlike the case of the 27 densely sampled ATCA epochs. Instead, we directly compare the measured flux densities across the epochs.

A comparison with archival LBA data reveals significant long-term variability. The 2010 archival observation shows a flux density of $160\pm24$~mJy, substantially higher than the flux measured during the first epoch of our monitoring in July 2024 ($68\pm10$~mJy; see Table~\ref{tab:LBA} and Figure \ref{fig:LBA_variability}). This indicates that the parsec-scale core has undergone pronounced variability on decade-long timescales, consistent with the known variable behaviour of NGC~7213 at radio and X-ray frequencies.

During our monitoring campaign, spanning approximately six months, we detect a clear increase in the LBA flux density, from $\sim70$~mJy to $\sim110$~mJy, corresponding to a net rise of $\sim40$~mJy. Given that the source remains unresolved throughout the campaign, this flux increase must originate within the compact core region probed by the LBA. The amplitude and timescale of the flux increase closely match those observed at 9~GHz with ATCA during the monitoring, over the same 6-month period, at a frequency very close to that of the LBA observations. This strongly suggests that the high-frequency radio variability detected by ATCA is dominated by the parsec-scale core rather than by more extended emission.

The consistency between the LBA and ATCA flux-density increases supports a scenario in which the radio variability is driven by changes in the innermost core/jet system. In the context of the recent X-ray flare observed in NGC~7213 \citep{Kammoun2025}, this behaviour is naturally interpreted as the radio counterpart of enhanced activity in the central engine, reinforcing the physical connection between the X-ray and radio-emitting regions in this low-luminosity AGN.

%%%%%%%%%%%%%%%%%%%%%%%%%%%%%%%%%%%%%%%%%%%%%%%%%%%%%%%%%%%
\section{Discovery of symmetric kiloparsec-scale radio knots}
\label{sec:radio_morphology}
The analysis of the MeerKAT 1.3\,GHz archival data reveals two unresolved radio knots with flux
densities of $\simeq 1.3$--$1.6$\,mJy, located $\sim 1\arcmin$ north and
south of the nucleus (see Fig. \ref{fig:MeerKAT}, projected distance $\simeq 5$\,kpc at 22\,Mpc;
\citealt{Hameed2001}).  
uGMRT observations at 300--700\,MHz confirm both components with nearly
identical flux densities ($\sim 1$\,mJy) and flat spectra
($\alpha \approx 0$, see Table\,\ref{table:hotspots_fluxes}), indicating compact synchrotron emission.

The two knots show a striking degree of symmetry.  
Their projected distances from the nucleus are nearly identical, their
flux densities agree within uncertainties, and the vectors
core–North and core–South differ in position angle by only
$\Delta{\rm PA} \simeq 6^\circ$ (conservatively $< 10^\circ$).  
Both knots also lie within $\sim 5''$ of the axis defined by the
core–opposite-knot direction.  
Such geometric and spectral symmetry strongly suggests a physical
connection with NGC\,7213.
The hypothesis that the two components are unrelated background sources
is therefore unlikely.  
A detailed estimate of the chance alignment probability, based on the
1.4\,GHz source counts of \citet{Richards2000}, source geometry and
spectral constraints, is presented in Sect.~\ref{sec:chance_alignment}.

\begin{figure}[]
   \centering
   \includegraphics[width=\hsize]{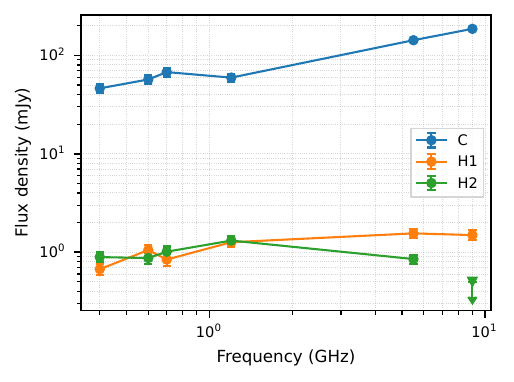}
      \caption{Radio spectra of the core (C) and the two hotspots (H1, H2) from 0.4 to 9 GHz. All components show a flat spectral index in the whole range.}
         \label{fig1}
   \end{figure}

\begin{figure*}
\sidecaption
    \includegraphics[width=12cm]{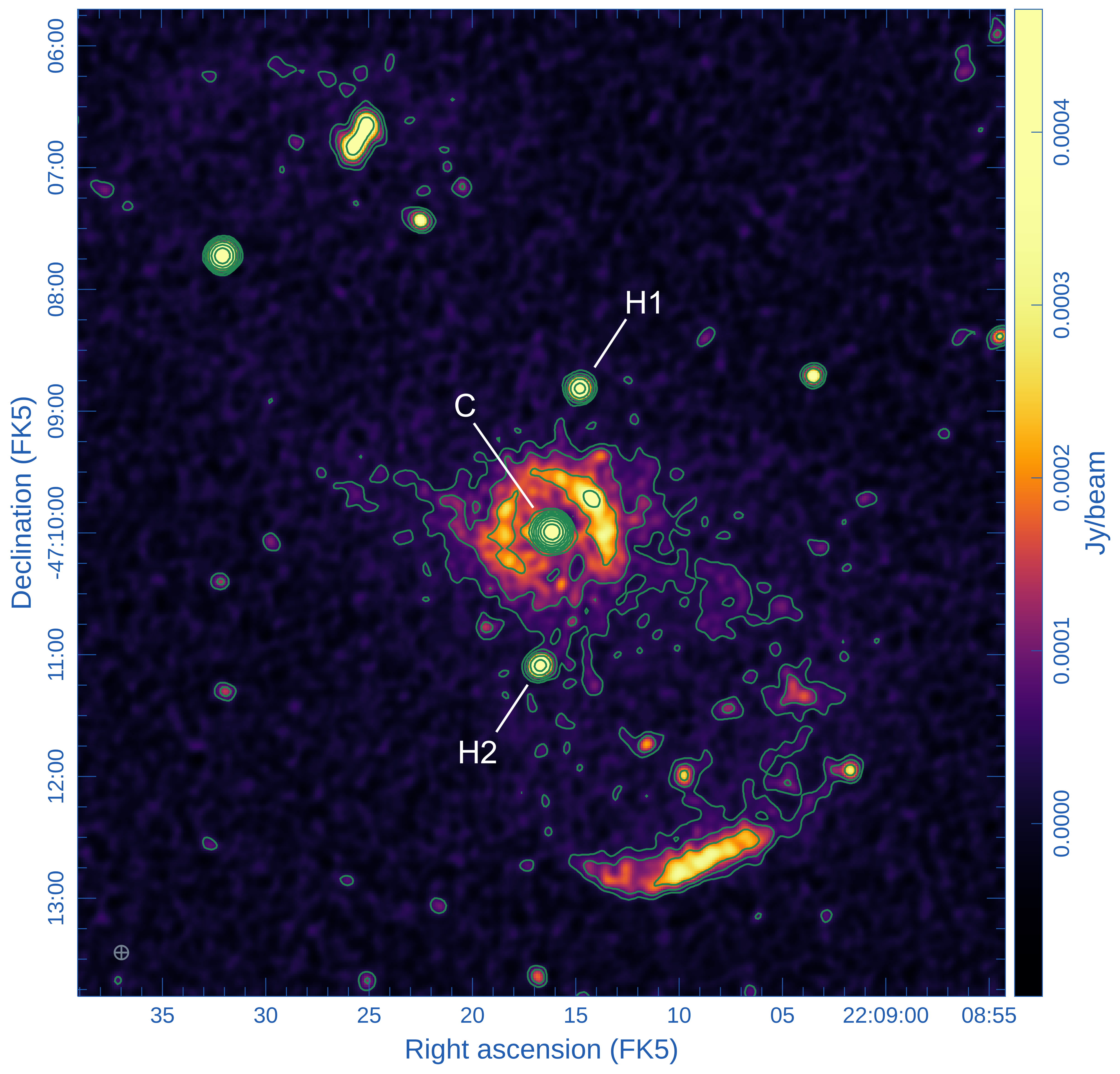}   % o il nome esatto del file
    \caption{MeerKAT 1.2\,GHz image of NGC\,7213. The central region shows circumnuclear ring of star formation of the host galaxy, the compact AGN core (C), and the northern (H1) and southern (H2) radio hotspots. Diffuse emission extending towards the south-east coincides with the H$\alpha$ filament reported by \citet{Hameed2001}. The coordinates for the hotspots are RA: $22{:}09{:}14.79\,(0.07)$, Dec: $-47{:}08{:}48.9\,(0.7)$ for H1, and RA: $22{:}09{:}16.71\,(0.07)$, Dec: $-47{:}11{:}05.4\,(0.7)$ for H2 (J2000). The synthesised beam is shown in the lower-left corner. Contours start at 3$\sigma$ and increase by a factor of two,
with $\sigma = 15\,\mu$Jy\,beam$^{-1}$.}  % copia il caption originale
    \label{fig:MeerKAT}
\end{figure*}

\begin{figure}
    \centering
    \includegraphics[width=1.0\linewidth]{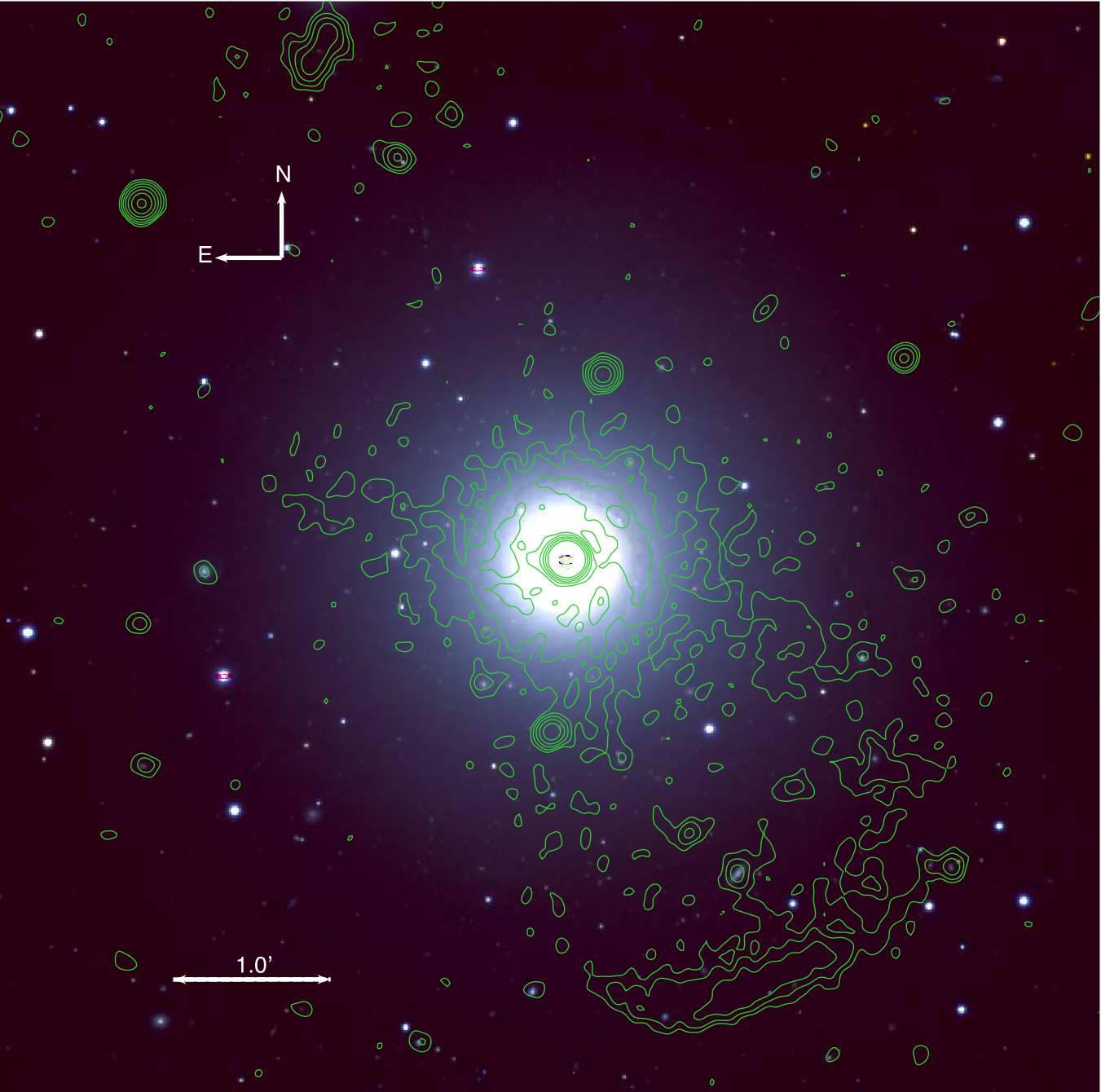}
    \caption{Superposition of the MeerKAT contours on an RGB composite optical image. The composite is made with images retrieved from the Legacy Survey DR10 in the \textit{z} (R), \textit{i} (G) and \textit{r} (B) bands. The diffuse emission from the host galaxy inner $\sim$5 kpc, surrounding the AGN, is visible, along with the H$\alpha$ filament to the south-west of the galaxy.}
    \label{fig:hotspots}
\end{figure}

ATCA observations provide further confirmation.  
Both knots are detected at 5.5\,GHz, preserving flat spectra up to this
frequency (see Fig. \ref{fig1}).  
The northern knot is also detected at 9\,GHz with a flat or slightly
declining spectrum, while the southern knot remains undetected but with
a $3\sigma$ upper limit ($S_9<1$\,mJy) fully consistent with a flat
spectrum.  
There is no evidence of any significant spectral asymmetry between the
two sides.
In order to capture the full spatial extent of the radio knots, we obtained a $5\arcmin\times5\arcmin$ Legacy Survey DR10 cutouts \citep{Dey_2019} in the \textit{r}, \textit{i} and \textit{z} bands to form a RGB composite. We show this composite in Figure \ref{fig:hotspots}, with superimposed MeerKAT contours. The location of the two knots falls within the radius of the host ($\sim5$ arcmin; \citealt{1989spce.book.....L})
The persistence of flat spectra up to 5.5--9\,GHz, combined with the
geometric symmetry of the knots, strongly favours an association with
NGC\,7213 and rules out a background-source interpretation (see also next section).

\begin{table*}[ht!]
\caption{Flux densities and spectral indices for the core (C) and hotspots (H1 and H2).} 
\label{table:hotspots_fluxes}
\centering
\begin{tabular}{ccccccccc}
\hline\hline
Component
& $S_{0.4}$
& $S_{0.6}$
& $S_{0.7}$
& $S_{1.2}$
& $S_{5.5}$
& $S_{9}$
& $\alpha_{0.4-1.2}$
& $\alpha_{1.2-5.5}$ \\
& (mJy)
& (mJy)
& (mJy)
& (mJy)
& (mJy)
& (mJy)
& 
&  \\
\hline
C
& $46.2 \pm 4.6$
& $56.9 \pm 5.7$
& $67.6 \pm 6.8$
& $59.4 \pm 5.9$
& $142.6 \pm 1.4$
& $186.1 \pm 1.8$
& $0.23 \pm 0.13$
& $0.58 \pm 0.07$ \\

H1
& $0.67 \pm 0.08$
& $1.05 \pm 0.13$
& $0.84 \pm 0.12$
& $1.26 \pm 0.13$
& $1.55 \pm 0.16$
& $1.49 \pm 0.17$
& $0.57 \pm 0.14$
& $0.14 \pm 0.10$ \\

H2
& $0.89 \pm 0.10$
& $0.87 \pm 0.11$
& $1.01 \pm 0.13$
& $1.31 \pm 0.13$
& $0.85 \pm 0.09$
& $<0.5$
& $0.35 \pm 0.14$
& $-0.28 \pm 0.10$ \\
\hline
\end{tabular}
\end{table*}

%%%%%%%%%%%%%%%%%%%%%%%%%%%%%%%%%%%%%%%%%%%%%%%%%%%%%%%%%%%
%%%%%%%%%%%%%%%%%%%%%%%%%%%%%%%%%%%%%%%%%%%%%%%%%%%%%%%%%%%
\subsection{Probability of chance alignment}
\label{sec:chance_alignment}

To quantify the likelihood that the two radio knots are unrelated
background sources, we use the 1.4\,GHz source counts from the recent
MIGHTEE survey \citep{Hale2023}, who find a surface density of
$\Sigma \simeq 3.5\times10^{-6}\,{\rm arcsec^{-2}}$ for sources brighter
than 1\,mJy (consistent with other deep MeerKAT fields; \citealt{Matthews2021}).

The probability of finding a single $>1$\,mJy background source within a
circular region of radius $5''$ is
\[
P_1 = \Sigma \pi (5'')^2 \simeq 2.75\times10^{-4}.
\]
The probability of finding two such sources in two distinct regions on
opposite sides of the nucleus is therefore
\[
P_2 = P_1^2 \simeq 7.56\times10^{-8}.
\]
Requiring that the two sources are aligned within $\pm10^\circ$ of the
jet axis introduces an additional geometric factor,
\[
f_{\rm align} \simeq 20^\circ/180^\circ \approx 0.11,
\]
yielding
\[
P_{\rm align} \simeq 8.4\times10^{-9}.
\]
Flat-spectrum sources constitute only $\sim10$--$15\%$ of the population
at this flux level \citep[e.g.][]{Richards2000,Smolcic2017}. Requiring
that both knots have flat spectra therefore adds a suppression
factor, $\sim(0.1)^2 \approx 10^{-2}$, giving a final probability of
\[
P_{\rm final} \sim 8 \times 10^{-11}.
\]
The background-source hypothesis is thus extremely unlikely.
The symmetry and spectra of the two knots strongly favour a physical
association with NGC\,7213.

We also searched for optical and infrared counterparts of the two radio
knots using the AllWISE catalogue \citep{Wright2010} and Gaia Data Release~3
\citep{GaiaDR3}. No Gaia DR3 source is found within 0.5\arcmin\ of either
H1 or H2. In the mid-infrared, no AllWISE source is present within
2\arcsec\ of the radio positions (i.e., $\sim$30\% of the MeerKAT beam, a more conservative threshold than the typical 10\% astrometric uncertainty).
The nearest AllWISE neighbours of H1 and H2 have mid-infrared colours
$W1-W2=-0.54$ and $-0.17$, respectively. These values are well below those
typically observed in AGN, which are characterised by red mid-infrared
colours ($W1-W2 \gtrsim 0.6$--0.8; \citealt{Stern2012,Assef2013}). The
nearby AllWISE sources are therefore unlikely to be associated with
background AGN and are considered unrelated to the radio knots.

We further inspected archival \textit{HST} imaging of NGC\,7213 and find no optical sources within 2\arcsec\ of the radio positions of either H1 or H2. The absence of compact optical counterparts at \textit{HST} resolution strengthens the conclusion that the radio knots are not associated with foreground stars or unrelated background galaxies.
The lack of plausible optical/IR counterparts further disfavours an interpretation in
terms of unrelated background AGN.

%%%%%%%%%%%%%%%%%%%%%%%%%%%%%%%%%%%%%%%%%%%%%%%%%%%%%
\subsection{Ruling out gravitational lensing by NGC\,7213}

Given the remarkable symmetry of the two radio knots detected at $\sim$5\,kpc
north and south of the nucleus, one might ask whether they could be multiple
images of a background source produced by gravitational lensing, with
NGC\,7213 acting as the lens. This scenario can be ruled out on quantitative
grounds.

At the distance of NGC\,7213 ($D_l \simeq 22$\,Mpc; \citealt{Hameed2001}),
$1$\,kpc corresponds to $\simeq 9\farcs4$, so the projected distance of each
knot from the nucleus is $\sim 5$\,kpc. Interpreting this separation as an
Einstein radius would require a lens mass of
$\sim 6\times10^{12}\,M_\odot$ within a projected radius of 5\,kpc
(for a singular isothermal sphere model).
Such a mass is typical of a massive galaxy cluster and is incompatible with
any reasonable mass model for an isolated Sa galaxy such as NGC\,7213, for
which no surrounding massive cluster is observed.

The probability of strong gravitational lensing producing two images with
$\Delta\theta \simeq 10''$ for a low-redshift source ($z \simeq 0.005$) is
extremely small, $P_{\rm lens} \lesssim 10^{-7}$ (using the lensing optical
depth from \citealt{Hilbert2007} and \citealt{Oguri2010}).
There is no evidence of lensed arcs or distorted background galaxies in deep
optical images of the field, and no obvious optical/IR counterparts that could
be identified as a single background AGN multiply imaged by NGC\,7213 (see
Sec.\ \ref{sec:chance_alignment}).

We therefore conclude that the symmetric radio knots cannot be explained as lensed images of a background source. Their geometry, spectra and location strongly favour a physical association with NGC\,7213, most naturally as the termination points (hotspots) of a weak or recurrent radio jet.
%%%%%%%%%%%%%%%%%%%%%%%%%%%%%%%%%%%%%%%%%%%%%%%%%%%%%%%%%%%%%%

%%%%%%%%%%%%%%%%%%%%%%%%%%%%%%%%%%%%%%%%%%%%%%%%%%%%%%%%%%%%%%

\section{Equipartition magnetic field in the hotspots}
We estimate the magnetic field strength in the two radio knots using the
revised equipartition and minimum-energy formalism of
\citet{BeckKrause2005}, which generalises the classical treatment of
\citet{Pacholczyk1970} and provides more robust estimates for compact
synchrotron sources with arbitrary spectral index and integration range.
This formulation explicitly accounts for the proton--electron energy
ratio $k$, the synchrotron spectral shape, and the integration limits
in frequency, and is widely adopted in modern analyses of radio jets,
LLAGN, and compact shock regions.

Each hotspot has a measured flux density of
$S_\nu \simeq 1.5$\,mJy at 1.3\,GHz and a flat spectrum between
300\,MHz and 1.3\,GHz.
Both knots are detected in the ATCA 5.5\,GHz image, preserving flat
spectra, and the northern knot is further detected at 9\,GHz.
The southern knot remains undetected at 9\,GHz, but its
$3\sigma$ upper limit ($S_9 < 1$\,mJy) is higher than the 5.5\,GHz flux,
indicating that both hotspots maintain flat spectra up to at least
5.5--9\,GHz.

The ATCA observations also provide stringent upper limits on the
intrinsic sizes of the two knots.
The northern hotspot (H1) is compact at 9\,GHz, with
$\theta_N < 0\farcs6$ ($r_N \lesssim 60$\,pc at 22\,Mpc).
For the southern hotspot (H2) we derive a looser constraint from the 5.5\,GHz
image, $\theta_S < 1\farcs23$ ($r_S \lesssim 120$\,pc).
Assuming spherical geometries, these correspond to
$V_{\rm H1} \lesssim 9\times10^{60}$\,cm$^{3}$ and
$V_{\rm H2} \lesssim 7\times10^{61}$\,cm$^{3}$.

We adopt a proton--electron energy ratio $k = 100$, a filling factor
$\phi = 1$, and integrate the synchrotron spectrum between
$\nu_1 = 10^{7}$\,Hz and $\nu_2 = 10^{11}$\,Hz, following
\citet{BeckKrause2005}.
The synchrotron luminosity of each hotspot is
\begin{equation}
L_{\rm rad} \approx 4\pi D^2 S_\nu (\nu_2 - \nu_1)
\simeq 9\times10^{37}\ {\rm erg\,s^{-1}}.
\end{equation}

Applying the revised equipartition formula yields lower limits on the
equipartition magnetic field strength of
\begin{equation}
B_{\rm H1,eq} \gtrsim 1.5 \times 10^{-4}\ {\rm G}, \qquad
B_{\rm H2,eq} \gtrsim 0.8 \times 10^{-4}\ {\rm G},
\end{equation}
where the quoted values correspond to the maximum volumes allowed by the
ATCA upper limits on the source size. These should be regarded as strict
lower limits, because the true emitting regions could be significantly
more compact than the current resolution limits.
The southern hotspot therefore has an equipartition field approximately
a factor of two smaller than the northern one, as expected from its
larger allowed volume.

The equipartition fields of
$B_{\rm eq} \sim (0.8\text{--}3)\times10^{-4}$\,G fall within the range
found in compact, shock-dominated regions of extragalactic radio jets.
Although these values overlap with those measured in the most compact
hotspots of powerful FR\,II radio galaxies 
\citep[e.g.][]{Meisenheimer1989,Hardcastle1998,Kataoka2005,Godfrey2013},
they do not imply FR\,II-like jet powers.  
As emphasised by \citet{BeckKrause2005}, equipartition fields scale
mainly with the ratio $L/V$, and compact termination regions can attain
$B \sim 10^{-4}$\,G independently of the global jet energetics.

Hotspots in FR\,I jets typically show lower fields
($10$--$50\,\mu$G; \citealt{LaingBridle2002,Hardcastle2002,Croston2008}),
but compact working surfaces embedded in dense galactic environments can
reach several $10^{-4}$\,G even in low-power jets.  
Comparable values are also observed in compact knots of LLAGN and
Seyfert jets \citep[e.g.][]{Kharb2016}, where
environmental confinement and strong shock compression dominate over 
intrinsic jet power.

The strong magnetic fields inferred for the two knots are fully
consistent with a weak or intermittent jet still propagating within the
dense, multiphase ISM of NGC\,7213.
At $\sim 5$\,kpc from the nucleus, the jet interacts with
H\,{\sc i} tidal debris and ionised filaments \citep{Hameed2001},
generating strong shocks that efficiently compress and amplify the
magnetic field.
Numerical relativistic hydrodynamic (RHD) simulations of low-power jets
in clumpy environments \citep[e.g.][]{Mukherjee2016,Mukherjee2018a}
routinely produce compact working surfaces with strong shocks.
The equipartition magnetic fields we derive 
($B_{\rm eq}\sim(0.8-3)\times10^{-4}$\,G) are consistent with the range 
expected in compact, shock-dominated regions of extragalactic jets, 
including those found in relativistic magnetohydrodynamic (RMHD) 
simulations of magnetised low-power jets 
\citep[e.g.][]{Mukherjee2020,Meenakshi2023}.

These magnetic-field values therefore trace the local
shock physics and environmental confinement rather than the global jet
power.
Together with their nearly identical spectra, symmetry, and lack of
extended lobes, the magnetic fields strongly support the interpretation
of the two knots as compact termination shocks of a weak or intermittent
jet still confined within the host galaxy of NGC\,7213.

%%%%%%%%%%%%%%%%%%%%%%%%%%%%%%%%%%%%%%%%%%%%%%%%%%%%%%%%%%%%%%%%%%%%%%%%%%%

\section{Large-scale morphology and the jet–environment connection}

The discovery of two compact radio knots located symmetrically at
$\sim$5\,kpc from the nucleus of NGC\,7213 provides new information on
the structure and environmental impact of this LLAGN.
These knots are embedded within a galaxy that already shows clear signs
of past dynamical disturbance, including a highly irregular H\,{\sc i}
distribution and a giant H$\alpha$ filament at 19\,kpc
\citep{Hameed2001}.  
Taken together, these features indicate that the central engine is
currently interacting with a dense and complex interstellar medium,
although they do not by themselves establish a recurrent jet history.
\subsection{Connection with the H$\alpha$ and H{\sc i} structures}

NGC\,7213 exhibits several signatures of past dynamical disturbance.
\citet{Hameed2001} discovered a 19\,kpc-long H$\alpha$ filament embedded
within a highly disturbed H\,{\sc i} tidal tail, extending to radii of
$\sim 60$\,kpc.  
The H\,{\sc i} morphology clearly indicates that NGC\,7213 is a merger
remnant, and the giant H$\alpha$ filament is naturally explained as part
of this tidal structure.  
Its ionisation mechanism remains ambiguous: photoionisation by UV
radiation escaping from the nucleus and shock ionisation associated with
local gas flows or a weak radio outflow were both considered plausible by
\citet{Hameed2001}, but the available data do not distinguish between
these possibilities.

The axis defined by the two symmetric radio knots is aligned to within
$\sim$10$^\circ$ of the direction of the H$\alpha$ filament.  
This geometric similarity may be coincidental, given the disturbed nature
of the system, but it suggests that the jet is propagating through a
medium shaped by past interactions and may encounter pre-existing
channels of lower density.  
The H$\alpha$/H\,{\sc i} filament therefore provides important context for
the jet–ISM interaction, even though it cannot be used as evidence of a past episode of jet activity.

%%%%%%%%%%%%%%%%%%%%%%%%%%%%%%%%%%%%%%%%%%%%%%%%%%%%%

\subsection{Evidence of a recurrent or intermittently powered jet}

The presence of two compact, symmetric radio knots at $\sim$5\,kpc,
combined with the absence of extended radio lobes even at low
frequencies, suggests that the jet in NGC\,7213 is relatively weak and
strongly influenced by its surrounding medium. 
The disturbed and multiphase H\,{\sc i} distribution surrounding the
galaxy provides a plausible physical context for such behaviour. 

Numerical simulations and recent reviews of jet--ISM interaction show
that jets propagating through clumpy, merger-disturbed gas can be
significantly affected by the ambient medium, leading to compact,
shock-dominated structures rather than large-scale lobes 
\citep[e.g.][]{Mukherjee2016,Mukherjee2018a,Mukherjee2025}. In these
environments, the jet evolution is expected to proceed through distinct
phases, including an initial `confined' stage in which the jet is
temporarily trapped within the dense ISM and propagates through
low-density channels (the so-called ``flood-and-channel'' mode),
inflating an over-pressurised cocoon before eventually breaking out 
\citep{Mukherjee2025}.

The efficiency of this interaction depends sensitively on the jet power
and on the structure of the ambient medium. While high-power jets tend
to rapidly drill through the ISM, low- and intermediate-power jets can
remain confined for longer timescales and experience stronger coupling
with the surrounding gas, transferring a significant fraction of their
kinetic energy into shocks and turbulence 
\citep{Wagner2012,Mukherjee2025}. This regime is particularly relevant
for low-accretion systems such as NGC\,7213. In particular,
\citet{Wagner2012} show that for $P_{\rm jet}/L_{\rm Edd}\lesssim10^{-4}$
the jet does not efficiently disrupt dense clouds but instead escapes
through porous low-density regions, resulting in localised and confined
feedback.

However, the observed symmetry and compactness of the knots do not 
constitute unambiguous evidence of strong jet--gas interaction. At projected distances of $\sim$5--10\,kpc the tidal debris are expected 
to have significantly lower densities than the central regions 
typically modelled in the simulations 
\citep[e.g.][]{Mukherjee2016,Mukherjee2018a,Mukherjee2025} 
(mean $n\sim100$\,cm$^{-3}$, dense cores $n>1000$\,cm$^{-3}$). In such lower-density environments a 
jet may propagate more ballistically, producing compact and symmetric 
working surfaces without requiring strong backflows or cocoon inflation 
\citep[e.g.][]{AntonuccioDelogu2010,Clarke1991}. Moreover, even in 
clumpy media the simulations frequently predict irregular or asymmetric 
morphologies due to local inhomogeneities 
\citep[e.g.][]{Mukherjee2018b,Gaibler2011}.

At the same time, it is important to note that the parameter space
explored by current simulations is not uniquely tailored to NGC\,7213.
For instance, the IC~5063 simulations of \citet{Mukherjee2018a} adopt
relativistic jets with $P_{\rm jet}\sim10^{44}$--$10^{45}\,\mathrm{erg\,s^{-1}}$
and $\Gamma\simeq4$--6 impacting a dense, clumpy disc. Although these
conditions are more extreme than those expected in NGC\,7213, several
key results—such as the development of compact working surfaces and
the importance of ISM porosity—are expected to scale primarily with the
ratio $P_{\rm jet}/L_{\rm Edd}$ and the gas distribution.
Finally, magnetohydrodynamic instabilities (e.g.\ Kelvin--Helmholtz or 
current-driven kink modes) can also affect jet propagation and hotspot 
symmetry, depending on jet power, magnetisation, and density contrast 
\citep{Mukherjee2020,Meenakshi2023}. 

Overall, the observations are consistent with a scenario in which a weak
jet is currently interacting with, and possibly confined by, the
disturbed ISM of NGC\,7213, although alternative interpretations 
(ballistic propagation through low-density channels or stable 
highly magnetised jets) cannot be excluded. The H$\alpha$/H\,{\sc i} 
filament provides evidence of a past merger, but does not by itself 
trace a previous phase of jet activity.
%%%%%%%%%%%%%%%%%%%%%%%%%%%%%%%%%%%%%%%%%%%%%%%%%%%%%
\subsection{Kinematic age of the hotspots}
\label{sec:kinematic_age}
The projected separation of each hotspot from the nucleus is
$R_{\rm proj} \simeq 5$\,kpc.
If the knots trace the current termination points of a bipolar jet,
a first-order estimate of the kinematic age can be obtained by assuming
a characteristic advance speed for the jet head.
For low- and intermediate-power jets, measurements based on spectral
ageing and dynamical modelling typically yield hotspot advance speeds
$v_{\rm hs} \sim 0.01$--$0.1\,c$
\citep[e.g.][]{Alexander2000,ODea2021,Hardcastle2013}.
Similar values are also obtained in numerical simulations of jets
propagating through a clumpy or multiphase ISM, where strong
jet--environment coupling can significantly slow down the advance of the
jet head \citep{Wagner2012,Mukherjee2016,Mukherjee2018a,Mukherjee2025}.

However, higher advance speeds may be expected in alternative scenarios,
such as more ballistic or recently restarted jets propagating through
lower-density regions, with values up to $\sim0.3$--$0.4\,c$ suggested by
numerical simulations and theoretical considerations
\citep[e.g.][]{Mukherjee2020}. Therefore, the characteristic advance
speed in NGC\,7213 is uncertain and may span a broader range depending on
the jet power and the properties of the surrounding medium.

The true deprojected distance of the hotspots is
$R = R_{\rm proj}/\sin\theta$, where $\theta$ is the angle between the
jet axis and the line of sight.
While $\theta$ cannot be directly constrained from the current data, the roughly
two-sided large-scale morphology and the lack of blazar-like behaviour in the
parsec-scale core argue against a jet axis closely aligned with the line of sight.
In particular, blazars---whose relativistically beamed jets point nearly towards
the observer---typically exhibit strong and rapid variability across multiple
wavelengths and on short timescales as a consequence of Doppler boosting and
jet dynamics \citep[e.g.][]{UrryPadovani1995}, but such extreme variability is
not observed in the LBA core of NGC\,7213.
Moreover, the nearly identical flux densities of the two knots are consistent
with only modest relativistic boosting effects on kiloparsec scales.
Adopting a conservative viewing angle $\theta \gtrsim 30^\circ$ implies
$R \lesssim 10$~kpc for each hotspot, keeping the system on genuinely
galactic scales rather than requiring tens-of-kiloparsec deprojected sizes.

The corresponding kinematic age is therefore
\begin{equation}
t_{\rm kin} \simeq \frac{R}{v_{\rm hs}}
               = \frac{R_{\rm proj}}{v_{\rm hs}\,\sin\theta}.
\end{equation}
For $R_{\rm proj}=5$\,kpc and $v_{\rm hs} = \beta c$ with
$\beta = 0.01$--$0.4$, we obtained
\begin{equation}
t_{\rm kin} \simeq 1.6\times 10^{6}
\left(\frac{\beta}{0.01}\right)^{-1}
\left(\frac{\sin\theta}{0.5}\right)^{-1}
\ {\rm yr},
\end{equation}
which corresponds to
\begin{equation}
t_{\rm kin} \sim 4\times10^{4} \text{--} 2\times10^{6}\ {\rm yr}
\label{eq:tkin}
\end{equation}
for plausible viewing angles $\theta \sim 30^\circ$--$60^\circ$ and the
full range of possible advance speeds. Thus, if the observed knots mark the current jet termination points, the
present episode of jet activity is likely to span $\sim10^{5}$--$10^{6}$\,yr,
although significantly shorter timescales (down to a few $\times10^{4}$\,yr)
cannot be excluded in the case of faster, more weakly interacting jet
propagation.

This estimate carries several caveats.
First, the hotspot advance speed in NGC\,7213 is unconstrained and may be
lower than $0.01\,c$ if the jet is strongly decelerated by the dense,
merger-disturbed ISM, which would increase the inferred age.
Conversely, faster propagation in a lower-density medium would reduce the
age accordingly.
Second, the knots may represent quasi-stationary shock regions rather
than ballistic features \citep[e.g.][]{Bicknell1994,Bicknell1995}, in
which case the kinematic age does not directly trace the duration of the
current activity cycle.
Nevertheless, the simple scaling above indicates that the structures we
observe are consistent with a relatively recent ($\lesssim$\,a few megayears)
phase of jet activity, compatible with duty cycles inferred for
LLAGNs \citep{BestHeckman2012}.

%%%%%%%%%%%%%%%%%%%%%%%%%%%%%%%%%%%%%%%%%%%%%%%%%%%%%%%%%%%%
\section{Discussion}
\label{sec:discussion}

The results presented in this work provide a coherent picture of radio activity in NGC~7213 across a wide range of spatial scales, from the sub-parsec core probed by the LBA to the kiloparsec-scale radio knots detected with MeerKAT, uGMRT, and ATCA. In this section we discuss the implications of these findings for the nature of the jet, its interaction with the surrounding interstellar medium, and its connection to the nuclear activity of this LLAGN.

\subsection{A confined weak jet from parsec to kiloparsec scales}

The LBA monitoring demonstrates that the radio core of NGC~7213 remains unresolved at all epochs, constraining the bulk of the 8~GHz emission to scales of $\lesssim0.2$--0.3~pc. Despite the absence of resolved structure, the core exhibits significant variability on both decade-long and month-long timescales. This behaviour is typical of compact radio cores associated with weak jets in LLAGN, where the radio emission is dominated by the innermost jet or jet-launching region rather than by extended, steady outflows. The ATCA monitoring at 5.5 and 9 GHz further supports this picture, showing that the flux increase is more pronounced at higher frequencies, consistent with variability originating in the innermost jet region and propagating to lower frequencies.

On much larger scales, the discovery of two compact radio knots located symmetrically at $\sim5$~kpc from the nucleus provides clear evidence that collimated outflows are able to propagate well beyond the central region of the galaxy. The coexistence of a variable, unresolved parsec-scale core and symmetric kiloparsec-scale knots naturally suggests that these structures are physically connected, tracing different stages of the same jet activity. In this picture, the LBA probes the current state of the jet close to the launching region, while the knots mark the locations where the jet energy is currently dissipated on galactic scales.

\subsection{Termination shocks and the role of the environment}

The flat radio spectra, compact sizes, and high equipartition magnetic fields inferred for the two knots strongly favour their interpretation as compact termination shocks or working surfaces rather than as extended radio lobes. Unlike classical FR\,II radio galaxies, where powerful jets inflate large lobes over hundreds of kiloparsecs, the jet in NGC~7213 appears unable to sustain such extended structures.

This behaviour is most naturally explained by environmental confinement. NGC~7213 is a merger remnant with a disturbed and multiphase interstellar medium, as traced by its highly irregular H\,{\sc i} distribution and the giant H$\alpha$ filament. Numerical simulations of weak jets propagating through clumpy, merger-disturbed media show that such conditions efficiently disrupt and decelerate jets, leading to the formation of compact, shock-dominated features with amplified magnetic fields, while preventing the development of large-scale lobes \citep[e.g.][]{Wagner2012,Mukherjee2016,Mukherjee2018a}. The properties of the knots in NGC~7213 closely match these expectations, indicating that local shock physics and environmental interaction dominate over the intrinsic jet power.

\subsection{Variability and connection to nuclear activity}

The increase of $\sim40$~mJy observed in the LBA flux density over a period of $\sim6$ months demonstrates that significant changes occur within the parsec-scale core on relatively short timescales. The fact that a comparable increase is observed at 9~GHz with ATCA over the same period indicates that the high-frequency radio variability is dominated by the compact core rather than by extended emission. In addition, the higher fractional variability measured at 9 GHz compared to 5.5 GHz indicates that the variability amplitude increases with frequency, as expected if the emission is dominated by progressively more compact regions closer to the jet base.

Such behaviour is consistent with previous studies of NGC~7213 showing correlated radio and X-ray variability, with radio flares lagging the X-rays by weeks \citep{Bell2011}. Although a detailed multi-wavelength timing analysis is beyond the scope of this paper, the LBA results provide strong evidence that the radio variability associated with the recent X-ray brightening originates in the innermost jet region. This supports a scenario in which changes in the accretion flow are rapidly communicated to the base of the jet, producing enhanced radio emission on parsec scales that may eventually feed larger-scale structures.

%%%%%%%%%%%%%%%%%%%%%%%%%%%%%%%%%%%%%%%%%%%%%%%%%%%%%%%%%

\subsection{Implications for jet duty cycles in LLAGN}
If the two hotspots represent the current termination points of a bipolar jet, their projected distance of $\sim5$\,kpc implies a kinematic age of order $10^{4}$--$10^{6}$\,yr for plausible hotspot advance speeds (Sect.~\ref{sec:kinematic_age}). This timescale is consistent with the short duty cycles inferred for low-luminosity AGN \citep{BestHeckman2012} and suggests that the present episode of jet activity is relatively recent.

At the same time, our LBA monitoring shows that the nuclear radio source remains unresolved down to sub-parsec scales at all epochs (Sect.~\ref{sec:lba_monitoring}), with no detectable extended jet structure connecting the core to the kiloparsec-scale hotspots despite significant variability on both month- and decade-long timescales. This non-detection is fully consistent with jet propagation in the flood-and-channel regime \citep{Mukherjee2018b,Mukherjee2025}, in which the relativistic plasma is efficiently channelled through low-density paths that act as natural waveguides. In this mode the jet head can maintain sufficient collimation to reach working surfaces at $\sim5$\,kpc while producing only faint or unresolved emission on parsec scales.

Combined with the absence of fossil radio lobes or extended low-frequency emission (even at uGMRT frequencies), the data indicate that either previous episodes were short-lived or that the jet power has always been low enough to remain strongly coupled to the dense, multiphase ISM. NGC\,7213 therefore provides an instructive nearby example of how weak jets in low-accretion systems can produce observable kiloparsec-scale structures without ever transitioning into a classical radio-galaxy phase, while the nuclear engine continues to exhibit recurrent activity on much shorter timescales.

%%%%%%%%%%%%%%%%%%%%%%%%%%%%%%%%%%%%%%%%%%%%%%%%%%%%%%%%%%%%

\section{Conclusions}

We summarise our main conclusions as follows:

\begin{itemize}

\item We discovered two compact radio knots located symmetrically at a projected distance of $\sim5$~kpc north and south of the nucleus of the LLAGN NGC~7213. The two components have nearly identical flux densities and flat radio spectra from 300~MHz up to at least 5.5~GHz.

\item ATCA observations confirm the flat spectral behaviour at higher frequencies. The northern knot is detected at 9~GHz, while the southern knot is consistent with a flat spectrum within upper limits. No significant spectral or geometric asymmetry is observed between the two sides.

\item Size constraints and equipartition arguments imply compact emitting regions with magnetic fields of $B\sim10^{-4}$~G, characteristic of shock-dominated working surfaces rather than extended radio lobes.

\item LBA monitoring shows that the radio core remains unresolved on sub-parsec scales at all epochs, but exhibits significant variability on both decade-long and month-long timescales, including a flux-density increase of $\sim40$~mJy over six months. The ATCA monitoring reveals significant variability at GHz frequencies, with a larger amplitude at 9 GHz than at 5.5 GHz, and a steady flux increase over the observing period. This behaviour, consistent with the LBA results, indicates that the variability is driven by the compact core and supports a scenario in which changes in the accretion flow propagate into the jet.

\end{itemize}

To conclude, the combined pc- and kpc-scale radio properties support a scenario in which NGC~7213 hosts a weak or intermittent jet that is strongly confined by a dense, merger-disturbed interstellar medium, producing compact termination shocks without inflating large-scale radio lobes.

%%%%%%%%%%%%%%%%%%%%%%%%%%%%%%%%%%%%%%%%%%%%%%%%%%%%%%%%%%%%%%
\section{Data availability}
Tables C.1 and D.1 are only available in electronic form at the CDS via anonymous ftp to \url{cdsarc.u-strasbg.fr} (130.79.128.5) or via \url{http://cdsweb.u-strasbg.fr/cgi-bin/qcat?J/A+A/}.
%%%%%%%%%%%%%%%%%%%%%%%%%%%%%%%%%%%%%%%%%%%%%%%%%%%%%%%%%%%%%%

\begin{acknowledgements}

FP acknowledges financial support from the Bando Ricerca Fondamentale INAF and "Programma
di Ricerca Fondamentale INAF 2023 and 2024.
A.L.T acknowledges support from ASI (Italian Space Agency) through Contract No. 2019-27-HH.0.

E.K. would like to thank the Science and Mission operation teams of the \textit{Swift} observatory for coordinating the observations.

R.R. acknowledges support from the European Research Council through 
the Consolidator grant BHianca (Grant agreement ID: 101002761).  

This paper makes use of data obtained with the MeerKAT radio telescope.
MeerKAT is operated by the South African Radio Astronomy Observatory
(SARAO), which is a facility of the National Research Foundation, an
agency of the Department of Science and Innovation of South Africa.

The uGMRT is operated by the National Centre for Radio Astrophysics of
the Tata Institute of Fundamental Research. We thank the staff of the
GMRT that made these observations possible.

The Australia Telescope Compact Array is part of the Australia
Telescope National Facility which is funded by the Australian
Government for operation as a National Facility managed by CSIRO.

The Long Baseline Array is part of the Australia Telescope National
Facility, operated by CSIRO. The LBA is funded by the Australian
Government for operation as a National Facility managed by CSIRO.

This research has made use of NASA’s Astrophysics Data System.

This work made use of data supplied by the UK Swift Science Data Centre at the University of Leicester.

The Legacy Surveys consist of three individual and complementary projects: the Dark Energy Camera Legacy Survey (DECaLS; Proposal ID \#2014B-0404; PIs: David Schlegel and Arjun Dey), the Beijing-Arizona Sky Survey (BASS; NOAO Prop. ID \#2015A-0801; PIs: Zhou Xu and Xiaohui Fan), and the Mayall z-band Legacy Survey (MzLS; Prop. ID \#2016A-0453; PI: Arjun Dey). DECaLS, BASS and MzLS together include data obtained, respectively, at the Blanco telescope, Cerro Tololo Inter-American Observatory, NSF’s NOIRLab; the Bok telescope, Steward Observatory, University of Arizona; and the Mayall telescope, Kitt Peak National Observatory, NOIRLab. Pipeline processing and analyses of the data were supported by NOIRLab and the Lawrence Berkeley National Laboratory (LBNL). The Legacy Surveys project is honored to be permitted to conduct astronomical research on Iolkam Du’ag (Kitt Peak), a mountain with particular significance to the Tohono O’odham Nation. NOIRLab is operated by the Association of Universities for Research in Astronomy (AURA) under a cooperative agreement with the National Science Foundation. LBNL is managed by the Regents of the University of California under contract to the U.S. Department of Energy. This paper used data obtained with the Dark Energy Camera (DECam), which was constructed by the Dark Energy Survey (DES) collaboration. Funding for the DES Projects has been provided by the U.S. Department of Energy, the U.S. National Science Foundation, the Ministry of Science and Education of Spain, the Science and Technology Facilities Council of the United Kingdom, the Higher Education Funding Council for England, the National Center for Supercomputing Applications at the University of Illinois at Urbana-Champaign, the Kavli Institute of Cosmological Physics at the University of Chicago, Center for Cosmology and Astro-Particle Physics at the Ohio State University, the Mitchell Institute for Fundamental Physics and Astronomy at Texas A\&M University, Financiadora de Estudos e Projetos, Fundacao Carlos Chagas Filho de Amparo, Financiadora de Estudos e Projetos, Fundacao Carlos Chagas Filho de Amparo a Pesquisa do Estado do Rio de Janeiro, Conselho Nacional de Desenvolvimento Cientifico e Tecnologico and the Ministerio da Ciencia, Tecnologia e Inovacao, the Deutsche Forschungsgemeinschaft and the Collaborating Institutions in the Dark Energy Survey. The Collaborating Institutions are Argonne National Laboratory, the University of California at Santa Cruz, the University of Cambridge, Centro de Investigaciones Energeticas, Medioambientales y Tecnologicas-Madrid, the University of Chicago, University College London, the DES-Brazil Consortium, the University of Edinburgh, the Eidgenossische Technische Hochschule (ETH) Zurich, Fermi National Accelerator Laboratory, the University of Illinois at Urbana-Champaign, the Institut de Ciencies de l’Espai (IEEC/CSIC), the Institut de Fisica d’Altes Energies, Lawrence Berkeley National Laboratory, the Ludwig Maximilians Universitat Munchen and the associated Excellence Cluster Universe, the University of Michigan, NSF’s NOIRLab, the University of Nottingham, the Ohio State University, the University of Pennsylvania, the University of Portsmouth, SLAC National Accelerator Laboratory, Stanford University, the University of Sussex, and Texas A\&M University. BASS is a key project of the Telescope Access Program (TAP), which has been funded by the National Astronomical Observatories of China, the Chinese Academy of Sciences (the Strategic Priority Research Program “The Emergence of Cosmological Structures” Grant \# XDB09000000), and the Special Fund for Astronomy from the Ministry of Finance. The BASS is also supported by the External Cooperation Program of Chinese Academy of Sciences (Grant \# 114A11KYSB20160057), and Chinese National Natural Science Foundation (Grant \# 12120101003, \# 11433005). The Legacy Survey team makes use of data products from the Near-Earth Object Wide-field Infrared Survey Explorer (NEOWISE), which is a project of the Jet Propulsion Laboratory/California Institute of Technology. NEOWISE is funded by the National Aeronautics and Space Administration. The Legacy Surveys imaging of the DESI footprint is supported by the Director, Office of Science, Office of High Energy Physics of the U.S. Department of Energy under Contract No. DE-AC02-05CH1123, by the National Energy Research Scientific Computing Center, a DOE Office of Science User Facility under the same contract; and by the U.S. National Science Foundation, Division of Astronomical Sciences under Contract No. AST-0950945 to NOAO.
\end{acknowledgements}

%%%%%%%%%%%%%%%%%%%%%%%%%%%%%%%%%%%%%%%%%%%%%%%%%%%%%%%%%%%%%%
\bibliographystyle{./aa.bst}   % o mnras, apj, ecc.
\bibliography{NGC7213_v3}
%%%%%%%%%%%%%%%%%%%%%%%%%%%%%%%%%%%%%%%%%%%%%%%%%%%%%%%%%%%%%%

\begin{appendix}
\onecolumn

\section{uGMRT images}
\label{app:uGMRT}

\begin{figure*}
    \centering
    \includegraphics[width=0.49\linewidth]{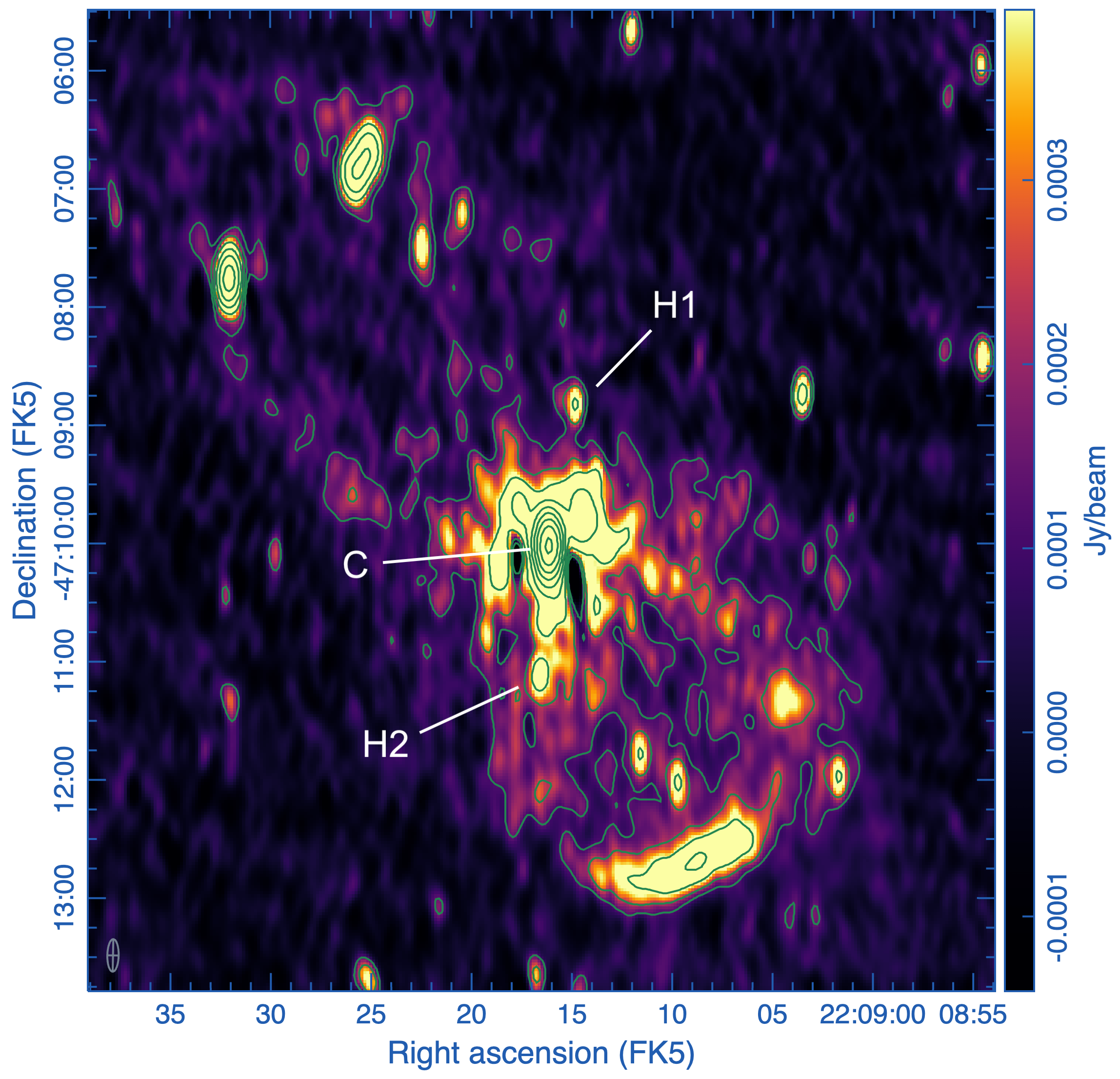}
    \includegraphics[width=0.49\linewidth]{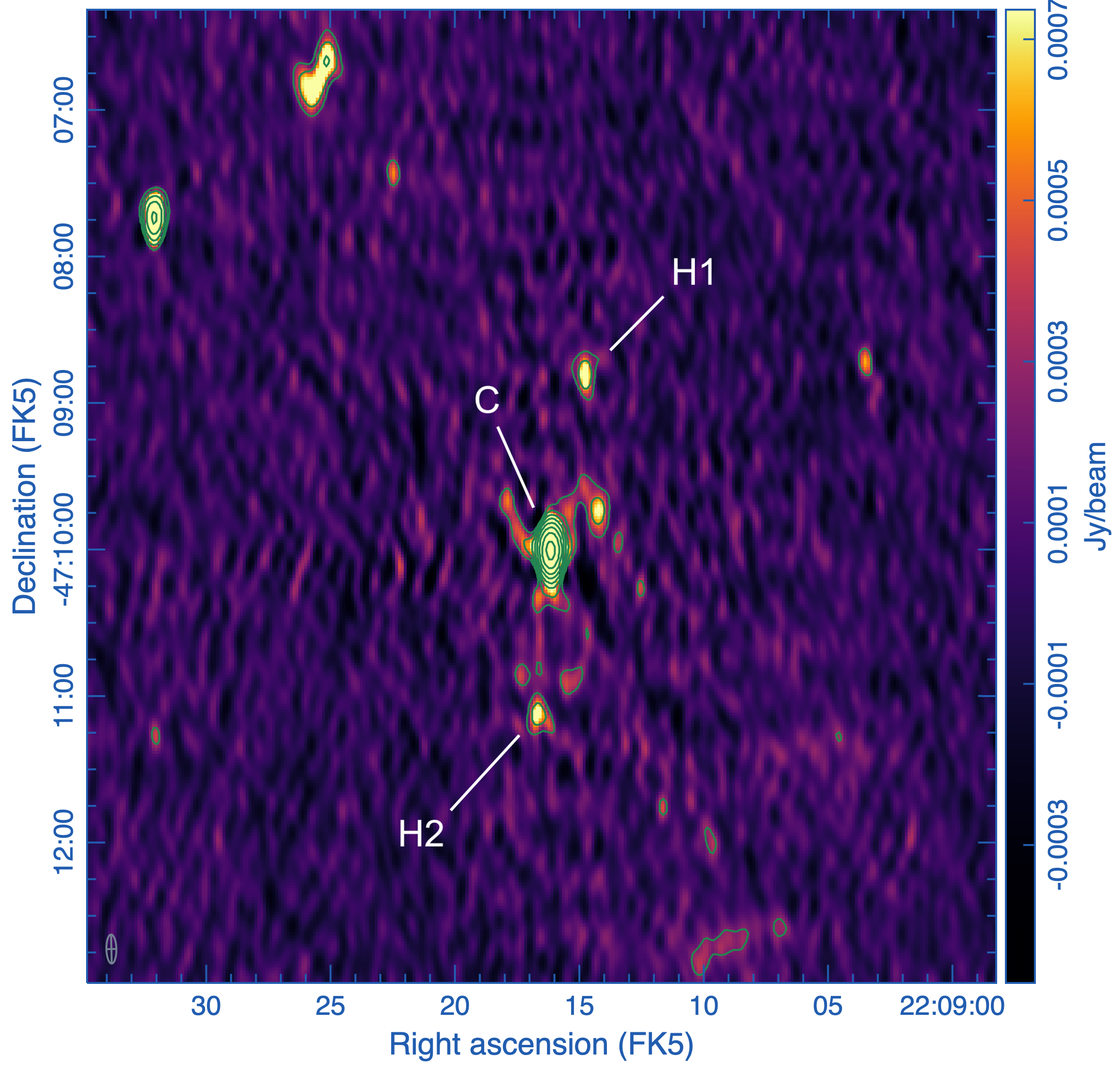}
    \caption{uGMRT images in Band-3 (0.4\,GHz, left) and Band-4 (0.6\,GHz, right). The compact AGN core (C), and the northern (H1) and southern (H2) radio hotspot are indicated. The synthesized beam is shown in the lower-left corner. Contours start at $3\sigma$ and increase by factors of two, with $\sigma=39\,\mu$Jy\,beam$^{-1}$ (Band-3) and $\sigma=80\,\mu$Jy\,beam$^{-1}$ (Band-4).}
    \label{fig:uGMRT}
\end{figure*}

The uGMRT images obtained in Band-3 (0.4\,GHz) and Band-4 (0.6\,GHz) are presented in Fig.~\ref{fig:uGMRT}. These observations were used to search for extended low-frequency emission and to derive the spectral indices of the compact knots (see Sect.~\ref{sec:radio_morphology}).

\begin{table*}[ht!]
\caption{Observation log of the uGMRT, MeerKAT and ATCA data used for imaging.}
\label{tab:uGMRT}
\centering
\begin{tabular}{lccccc}
\hline\hline
Telescope & Band & Freq. (GHz) & Date (dd-mm-yyyy) & FWHM (arcsec) & RMS ($\mu$Jy beam$^{-1}$) \\
\hline
uGMRT & B3 & 0.40 & 07/11/2025 & $16.5\times5.7$ & 39 \\
uGMRT & B4 & 0.65 & 24/11/2025 & $10.7\times3.6$ & 80 \\
MeerKAT & L & 1.28 & 03/01/2023 & $6.7\times6.7$ & 15 \\
ATCA & C & 5.5 & 30/07/2024 & $15.6\times1.2$ & 35 \\
ATCA & X & 9.0 & 03/09/2024 & $5.3\times0.8$ & 100 \\
\hline
\end{tabular}
\end{table*}

\section{LBA monitoring}
\label{app:LBA}

Details of the three LBA observing epochs (including the archival 2010 observation) and the measured 8\,GHz flux densities are reported in Table~\ref{tab:LBA}. These data were used for the parsec-scale variability analysis presented in Sect.~\ref{sec:lba_monitoring}.

\begin{table}[h!]
\caption{LBA observing epochs and measured 8\,GHz flux densities.}
\label{tab:LBA}
\centering
\begin{tabular}{cccc}
\hline\hline
Epoch & Date & $S_{8\,\rm GHz}$ (mJy) & Beam (mas) \\
\hline
Archive & 2010-03-11 & $160 \pm 24$ & $4.3\times3.9$ \\
1 & 2024-07-25 & $68 \pm 10$ & $2.2\times1.3$ \\
2 & 2024-11-04 & $106 \pm 16$ & $3.1\times1.9$ \\
3 & 2025-02-08 & $110 \pm 17$ & $2.0\times1.3$ \\
\hline
\end{tabular}
\end{table}

\section{ATCA monitoring}
\label{app:ATCA}

The flux densities measured during the 27-epoch ATCA monitoring campaign at 5.5 and 9.0\,GHz are listed in Table C.1 (available at the CDS). Uncertainties include both the statistical image noise and a conservative 5\% flux-scale calibration error. These data were used for the variability analysis and the computation of $F_{\rm var}$ and $R_X$ presented in Sect.~\ref{sec:atca_swift}.

\section{Swift/XRT monitoring}
\label{app:Swift}

The best-fit parameters of the \textit{Swift}/XRT spectra (2--10\,keV), modelled with a power-law plus Galactic absorption ($N_{\rm H}=1.08\times10^{20}$\,cm$^{-2}$; \citealt{HI4PI}), are reported in Table D.1 (available at the CDS). All spectra were fitted using Cash statistics \citep{Cash1979}. These results were used to construct the X-ray light curve and to compute the radio-loudness parameter $R_X$ (Sect.~\ref{sec:atca_swift}).

\end{appendix}
%%%%%%%%%%%%%%%%%%%%%%%%%%%%%%%%%%%%%%%%%%%%%%%%%%%%%%%%%%%%%%
\end{document}